\documentclass[aps,prl,twocolumn,showpacs,groupedaddress,nofootinbib]{revtex4}  
\usepackage{graphicx}  
\usepackage{dcolumn}   
\usepackage{bm}        
\usepackage{amsmath}   
\usepackage{amssymb}   
\usepackage{url}
\usepackage{xspace}
\usepackage{color}

\newcommand{\dzero}{D0\xspace}

\newcommand{\ttbar}{\ensuremath{t\bar{t}}\xspace}
\newcommand{\wjets}{\ensuremath{W}+\rm jets\xspace}
\newcommand{\pythia}{{\sc pythia}\xspace}
\newcommand{\vecbos}{{\sc vecbos}\xspace}
\newcommand{\ptmiss}{\ensuremath{{p\kern-0.5em\slash}_{T}}\xspace}

\begin{document}
\hspace{5.2in} \mbox{FERMILAB-PUB-09-285-E}

\title{Direct measurement of the mass difference between top and antitop quarks}
%
\author{V.M.~Abazov$^{37}$}
\author{B.~Abbott$^{75}$}
\author{M.~Abolins$^{65}$}
\author{B.S.~Acharya$^{30}$}
\author{M.~Adams$^{51}$}
\author{T.~Adams$^{49}$}
\author{E.~Aguilo$^{6}$}
\author{M.~Ahsan$^{59}$}
\author{G.D.~Alexeev$^{37}$}
\author{G.~Alkhazov$^{41}$}
% ...

\affiliation{\vspace{0.1 in}(The D\O\ Collaboration)\vspace{0.1 in}}
% ...
\affiliation{$^{6}$University of Alberta, Edmonton, Alberta, Canada;
                Simon Fraser University, Burnaby, British Columbia, Canada;
                York University, Toronto, Ontario, Canada and
                McGill University, Montreal, Quebec, Canada}
% ...
\affiliation{$^{30}$Tata Institute of Fundamental Research, Mumbai, India}
% ...
\affiliation{$^{37}$Joint Institute for Nuclear Research, Dubna, Russia}
% ...
\affiliation{$^{41}$Petersburg Nuclear Physics Institute,
                St. Petersburg, Russia}
% ...
\affiliation{$^{49}$Florida State University, Tallahassee, Florida 32306, USA}
% ...
\affiliation{$^{51}$University of Illinois at Chicago,
                Chicago, Illinois 60607, USA}
% ...
\affiliation{$^{59}$Kansas State University, Manhattan, Kansas 66506, USA}
% ...
\affiliation{$^{65}$Michigan State University,
                East Lansing, Michigan 48824, USA}
% ...
\affiliation{$^{75}$University of Oklahoma, Norman, Oklahoma 73019, USA}
% ...

\date{September 26, 2009}

\begin{abstract}We present a measurement of the mass difference  between $t$ and
$\overline{t}$ quarks in lepton+jets final states of $t\overline{t}$
events in 1 fb$^{-1}$ of data collected with the \dzero detector from Fermilab
Tevatron Collider $p\bar{p}$ collisions at $\sqrt{s}=1.96$~TeV. The measured
mass difference of $3.8\pm3.7$ GeV is consistent with the equality of $t$ and
$\overline{t}$ masses. This is the first direct measurement  of a mass
difference between a quark and its antiquark partner. \end{abstract}

\maketitle 

The $CPT$ theorem~\cite{Schwinger et al}, which is fundamental to any local
Lorentz-invariant quantum field theory, requires that the mass of a particle and
that of its antiparticle be identical. Tests of $CPT$ invariance for many of the
elementary particles accommodated within the standard model (SM) are available
in the literature~\cite{PDG}. Despite the fact that no violations have ever been
observed, it is important  to search for the possibility of $CPT$ violation in all
sectors of the standard model. Because quarks carry color, they cannot be
observed directly, but must first evolve through quantum chromodynamic (QCD)
interactions into jets  of colorless particles. These jet remnants reflect the
characteristics of the initially  produced quarks, such as their charges, spin
states, and masses. If the lifetimes of quarks are much longer  than the time
scale for QCD processes, the quarks form hadrons before they emerge  from
collisions, and decay from within bound hadronic states. This makes it
difficult  to measure a $q - \overline{q}$ mass difference because of the model
dependence of QCD binding and evolution processes. However, since the lifetime
of the top quark is far shorter than the time scale for QCD interactions, the
top-quark sector provides a way to measure the mass difference less
ambiguously~\cite{Cembr}. 

In this Letter, we report a measurement of the difference between the mass of
the top quark ($t$) and that of its antiparticle ($\overline{t}$) produced in
$p\overline{p}$ collisions at $\sqrt{s}=1.96$~TeV. Our measurement is based on
data corresponding to $\sim$1 fb$^{-1}$  of integrated luminosity collected with
the \dzero detector~\cite{Det} during Run II of the Fermilab Tevatron Collider. 
The events used in this analysis, identical to those in Ref.~\cite{P17PRL}, are
top quark pair (\ttbar) events in the lepton $+$ jets channel ($\ell+$jets) where
each top quark is assumed to always decay into a $W$ boson and a $b$ quark. One
of the $W$ bosons decays via $W\rightarrow \ell\nu$ into two leptons, and the other
one through $W\rightarrow q\overline{q}^{\prime}$ into two quarks, and all four
quarks ($q\overline{q}^{\prime}b\overline{b}$) evolve into jets.

We select events having one isolated electron (muon) with transverse momentum
$p_{T}>20$~GeV and $\left|\eta\right|<1.1$ ($\left|\eta\right|<2$), missing
transverse momentum $\ptmiss>20$~GeV, and exactly four jets with $p_{T}>20$~GeV
and $\left|\eta\right|<2.5$, where the pseudorapidity
$\eta=-\ln\left[\tan(\theta/2)\right]$, and $\theta$ is the polar angle with
respect to the proton beam direction. At least one of the jets is required to be
identified as a $b$-jet candidate.  A minimum azimuthal separation is required between
lepton $p_T$ and $\ptmiss$ vectors to further reduce multijet background arising
from lepton or jet energy mismeasurements.  The positively (negatively) charged
leptons are used to tag the $t$ ($\overline{t}$) in each event.  To reduce
instrumental effects that can cause charge dependent asymmetries in lepton
energy scale and resolution, solenoid and toroid magnetic field polarities are
routinely reversed.

The selected data sample consists of 110 $e+$jets and 110 $\mu$+jets events. 
The $W^+$ ($W^-$) boson decays into hadrons in 105 (115) events and into leptons
in 115 (105) events, consistent with invariance under charge conjugation.  The
fraction of \ttbar events in this sample is estimated to be $74\%$. The
background consists of \wjets and multijet events, with the latter comprising
$12\%$ of the entire background.

This analysis uses the matrix element (ME) method which relies on the extraction
of the properties of the top  quark (e.g., the mass) through a likelihood
technique based on  probability densities (PD) for each event, calculated from
the ME for the two major processes ($t\overline{ t}$ and \wjets production) that
contribute to the selected $\ell+$jets sample.  In calculating the PD for
$t\overline{t}$ production, we include only the leading order (LO) ME from 
$q\overline{q} \rightarrow \ttbar$ production~\cite{P14PRD}. We assume SM-like
\ttbar production and decay, where identical particle and antiparticle masses
are assumed for $b$ quarks and $W$ bosons but not for top quarks.  For \wjets
production, we use the ME provided in \vecbos~\cite{VECBOS}.  The PD for each
event is given in terms of the  fraction of  signal ($f$) and of background
($1-f$) in the data and the masses of the $t$ ($m_t$) and the $\overline{t}$
($m_{\overline{t}}$):
\begin{equation}
P_{\rm evt}=A(x)[fP_{\rm sig}(x;\,m_t,m_{\overline{t}})+(1-f)P_{\rm bkg}(x)]\ ,
\label{eq:MEpevt}
\end{equation}
where $x$ denotes the measured jet and lepton energies and angles, $A(x)$ is a
function only of $x$ and accounts for the geometrical acceptance and
efficiencies, and  $P_{\rm sig}$ and $P_{\rm bkg}$ represent the PD for \ttbar
and \wjets production, respectively.  Multijet events are also represented by
$P_{\rm bkg}$ since $P_{\rm bkg}\gg P_{\rm sig}$ for such events~\cite{pbkgqcd}.

The free parameters in Eq.~\ref{eq:MEpevt} are  determined from a likelihood
$L(x;\,m_t,m_{\overline{t}},f)$ constructed from the product of the $P_{\rm
evt}$ for all events. Jet energies are scaled by an overall jet energy scale
(JES) calibration factor derived by constraining the reconstructed mass of the
two jets from $W\rightarrow q\overline{q}^{\prime}$ decays in \ttbar events to
$80.4$~GeV~\cite{P17PRL,PDG}. The likelihood is maximized as a function of $f$
for each ($m_t$,$m_{\overline{t}}$) hypothesis to determine $f^{\rm best}$. An
integration of  the likelihood for $f=f^{\rm best}$ over the sum $m_{\rm
sum}=(m_t+m_{\overline{t}})/2$  results in a one-dimensional likelihood
$L(x;\Delta)$ as a function of mass difference  $\Delta=m_t - m_{\overline{t}}$.
This is used to extract the mean value of $\Delta$  and its uncertainty. A
similar procedure involving an integration over $\Delta$ gives $L(x;m_{\rm
sum})$ which is used to extract the mean value of $m_{\rm sum}$ and its
uncertainty.

The variables in any ME refer to nascent produced particles (leptons and
partons), but the measured quantities correspond to physical leptons and jets.
This difference is taken into account in the calculation of the event
probability by  convoluting over phase space a transfer function, $W(y,x)$, that
provides the resolution for  the lepton in question or a mapping of the observed
jet variables in an event ($x$)  to their progenitor parton  variables ($y$):
\begin{eqnarray}
P_{\textrm{sig}} & = & \frac{1}{\sigma_{\textrm{norm}}^{t\overline{t}}}\nonumber\\
& & \times\int{\displaystyle \sum}
d\sigma(y;m_t,m_{\overline{t}})dq_{1}dq_{2}F(q_{1})F(q_{2})W(y,x),\nonumber\\
\label{eq:MEpsgn}
\end{eqnarray}
where $d\sigma(y;m_t,m_{\overline{t}})$ is the leading-order partonic
differential  cross section, $q_{1}$ and $q_{2}$ are the momentum fractions of
the colliding  partons (assumed to be massless)  within the incident $p$ and
$\overline{p}$,  and the sum runs over all possible combinations of
initial-state parton flavors, jet-to-parton assignments, and all
$W\rightarrow\ell\nu$ neutrino solutions~\cite{nusol}. In the sum over
jet-to-parton assignments in $P_{\rm sig}$, each permutation  of jets carries a
weight $w_{i}$, which is the normalized product of probabilities
for tagging any jet under a given parton flavor hypothesis~\cite{P17PRL}. The
$F(q_i)$ include the probability densities for finding a parton of given flavor
and longitudinal momentum fraction in the $p$ or $\overline{p}$ assuming the
CTEQ6L1~\citep{cteq} parton distribution functions (PDF), as well as the
probability densities  for the transverse components of the $q_i$ obtained from
the LO event generator \pythia~\citep{pythia}. The normalization term 
$\sigma_{\textrm{norm}}^{\ttbar}$ is described below.
\begin{figure}
\begin{centering}
\includegraphics[width=0.5\columnwidth]{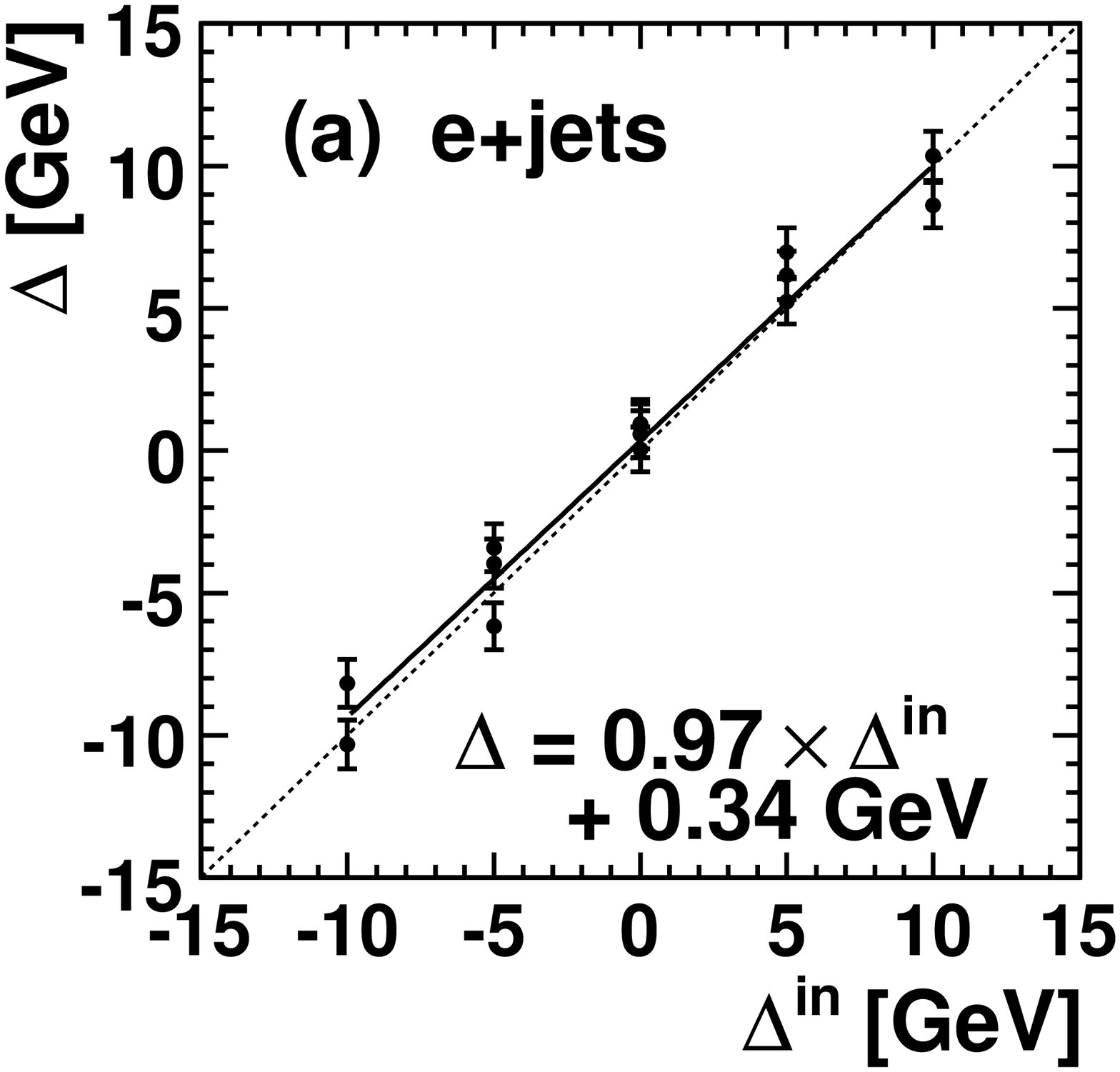}\includegraphics[width=0.5\columnwidth]{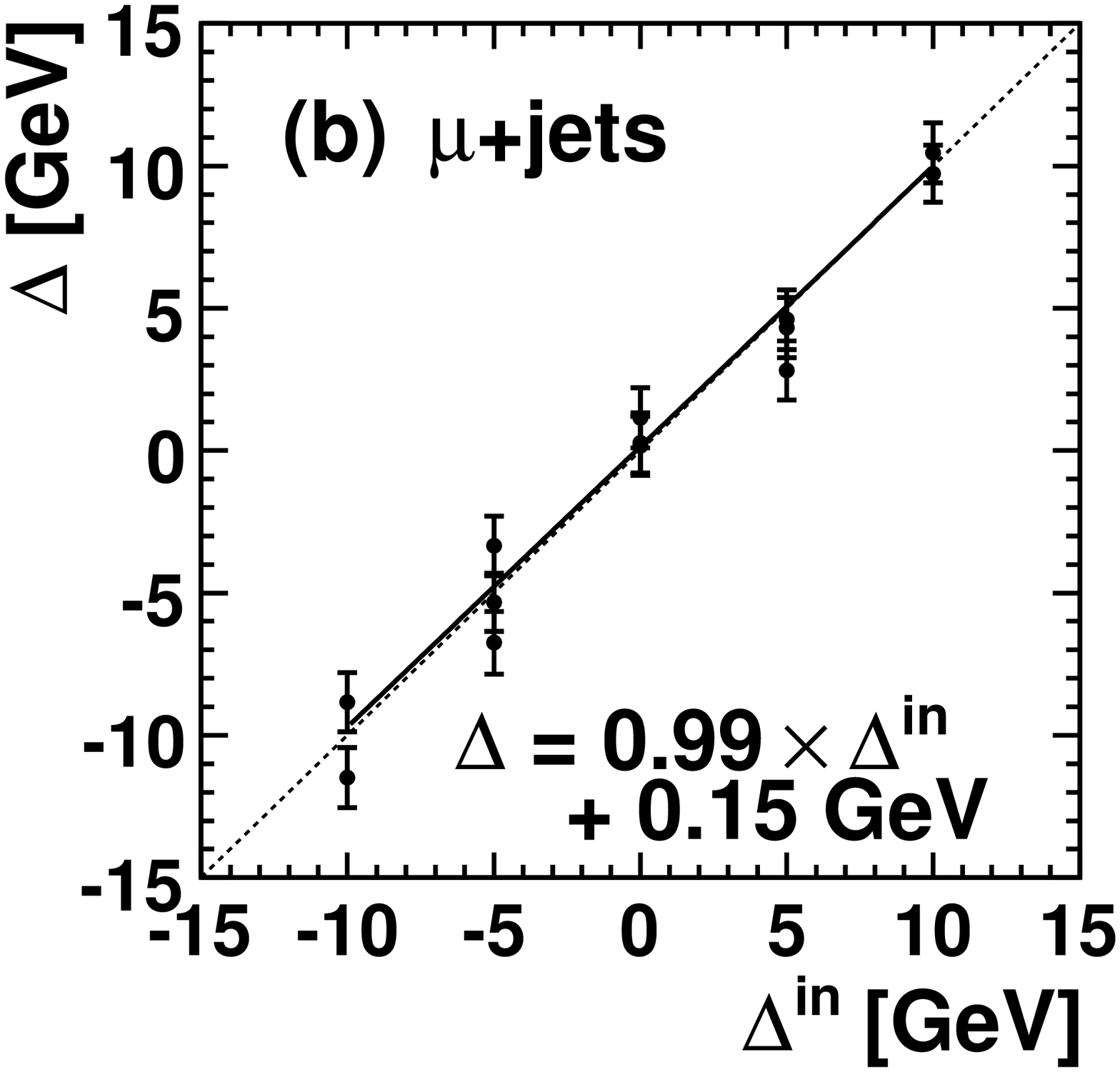} 
\par\end{centering}
\caption{\label{fig:calib}Values of the measured mean $\Delta$ from MC pseudo
experiments as a function of $\Delta^{\rm in}$, parameterized by straight lines
for (a) $e+$jets and (b) $\mu+$jets MC events. Dotted lines represent complete
equality between measured and input values. Results from pseudo experiments with
same $\Delta^{\rm in}$ but different $m_{\rm sum}$ correspond to the extra
points for fixed $\Delta^{\rm in}$ (see text).}
\end{figure}

The overall detection efficiency for \ttbar depends on the values of both $m_t$
and $m_{\overline{t}}$. This is taken into account through the normalization by
the observed cross section  $\sigma_{\textrm{norm}}^{\ttbar}=\smallint
A(x)P_{\textrm{sig}}dx= \sigma^{t\overline{t}}(m_t,m_{\overline{t}})\left\langle
A(m_t,m_{\overline{t}})\right\rangle$, where
$\sigma^{t\overline{t}}(m_t,m_{\overline{t}})$ is the total cross section
calculated by integrating the partonic cross section
$\sigma^{t\overline{t}}_{q\overline{q}}$~\cite{sigtot}, corresponding to the
specific ME used in the analysis, over initial and final parton distributions
and summing over initial parton flavors. $\left\langle
A(m_t,m_{\overline{t}})\right\rangle$ is the mean acceptance determined from the
generated $t\overline{t}$ events. The expressions for $P_{\rm bkg}$ are similar,
except that the probability does not depend on $m_t$ or $m_{\overline{t}}$.

Samples of \ttbar MC events with different values of  $m_t$ and
$m_{\overline{t}}$ are required to simulate \ttbar production and decay in order
to calibrate the results of the analysis. These events are generated with a
version of the \pythia generator~\cite{pythia} modified to provide independent
values of $m_t$ and $m_{\overline{t}}$. The specific values  chosen for ($m_t$,
$m_{\overline{t}}$) form a square grid spaced at 5 GeV  intervals between
(165,165) and (180,180), excluding the two extreme points at  (165,180) and
(180,165). The MC events for equal values of $m_t$ and $m_{\overline{t}}$ are
generated with the default version of \pythia. 

Approximations made in formulating the likelihood can bias the final result.
This issue is examined by comparing the measured and input values of $\Delta$ 
in pseudo experiments composed of MC \ttbar and \wjets events. The calibration
is shown in Fig.~\ref{fig:calib} in terms of the measured mean $\Delta$ as a
function of its input value ($\Delta^{\rm in}$), separately for  the $e+$jets
and $\mu+$jets MC samples, for all MC samples generated at the input reference
points on the ($m_t$, $m_{\overline{t}}$) grid.  There are 2, 3, 4, 3, and 2
different ($m_t$, $m_{\overline{t}}$) points with a common $\Delta^{\rm in}$ of
$-10$, $-5$, 0, $+5$, and $+10$ GeV, respectively.  The dispersions in the
measured values of mean $\Delta$ for different ($m_t$, $m_{\overline{t}}$)
points, but with same values of $\Delta^{\rm in}$, are consistent with expected
statistical fluctuations, as can be observed in Fig.~\ref{fig:calib}. The fit 
$\chi^2$/d.o.f. for the points in Figs.~\ref{fig:calib}(a) and
\ref{fig:calib}(b) are 1.8 and 0.84, respectively. The parameterizations shown
in Fig.~\ref{fig:calib} are used to calibrate $L(x;\Delta)$ for the selected
data sample.

We define the pull as $(\Delta-\langle\Delta\rangle)/\sigma(\Delta)$ where
$\Delta$ is the measured mass difference for a given pseudo experiment,
$\langle\Delta\rangle$ is the mean measured mass difference for all pseudo
experiments, and $\sigma(\Delta)$ is the uncertainty of the measured mass
difference for the given pseudo experiment.  The mean widths of the pull
distributions for all samples used in Fig.~\ref{fig:calib} are 1.2 and 1.1 for
$e+$jets and $\mu+$jets, respectively.  The deviations of these widths from 1
are used to correct the measured uncertainties in data.
\begin{figure}
\begin{centering}
\includegraphics[width=0.5\columnwidth]{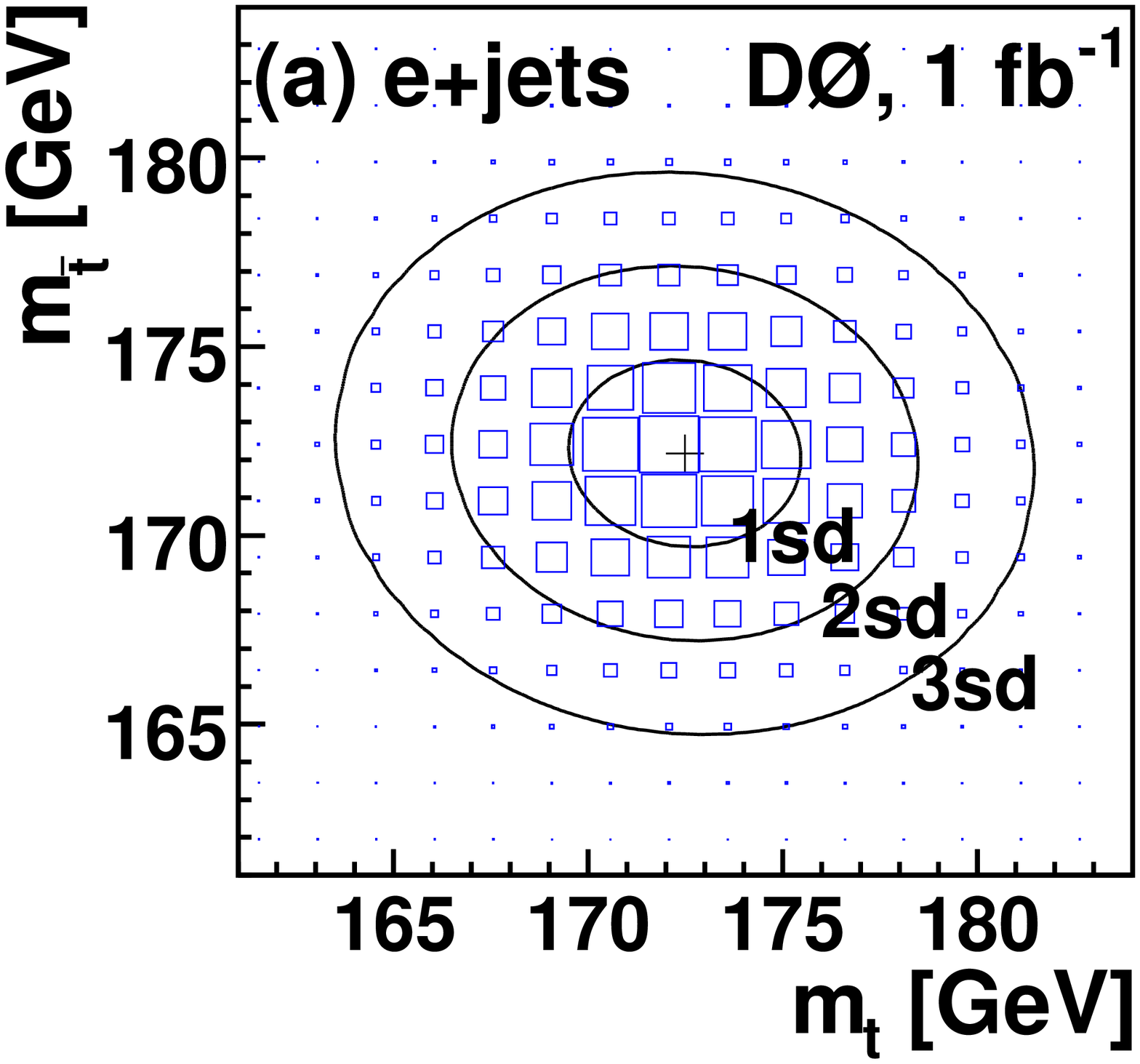}\includegraphics[width=0.5\columnwidth]{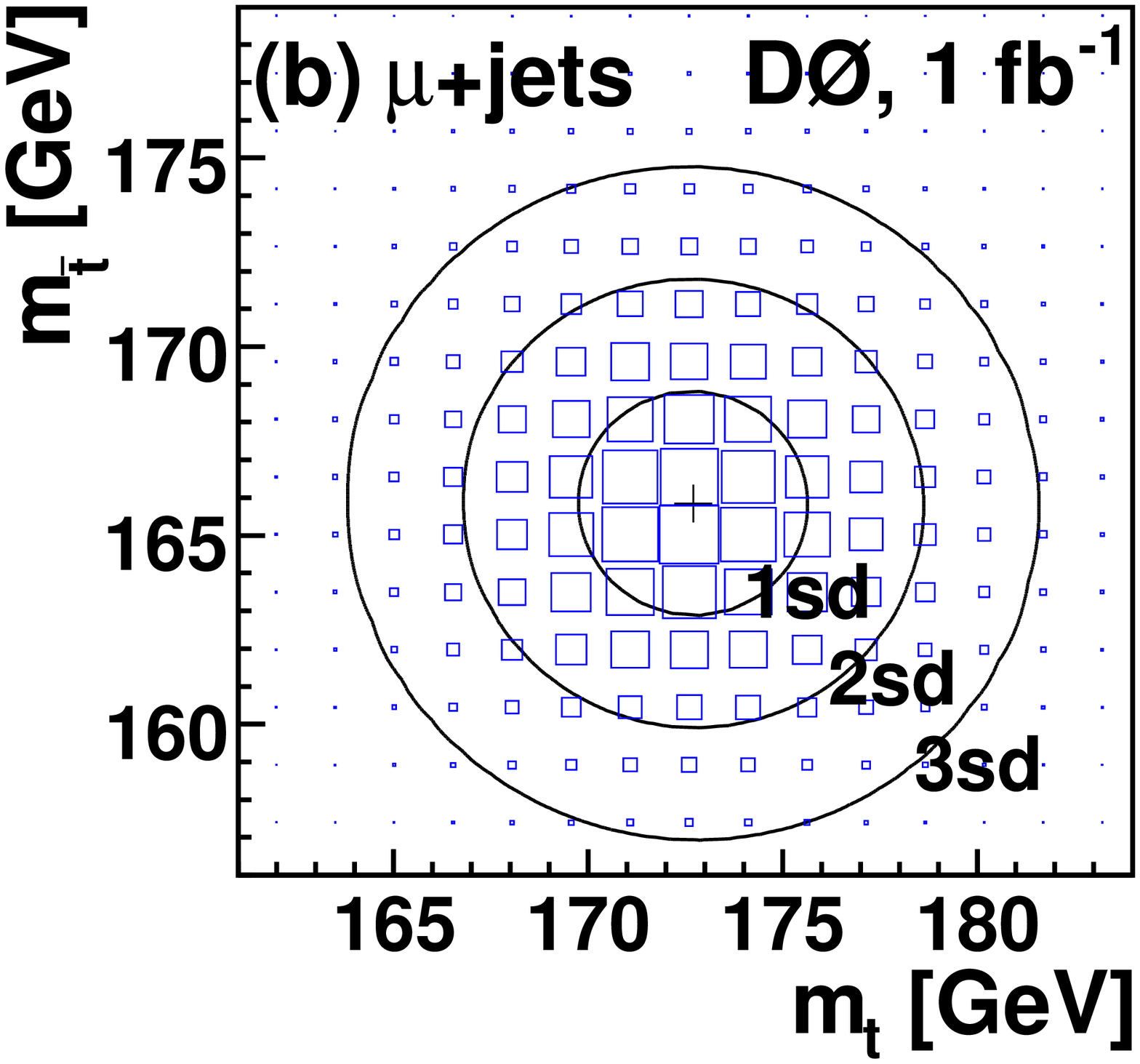} 
\par\end{centering}
\caption{\label{fig:2d}Fitted contours of equal probability for the
two-dimensional likelihoods as a function of $m_t$ and $m_{\overline{t}}$ for
(a) $e+$jets  and (b) $\mu+$jets data.  The boxes, representing the bins in the
two-dimensional histograms of the likelihoods, have areas proportional to the
bin contents, set equal to the value of the likelihood evaluated at the bin
center.}
\end{figure}

\begin{figure}
\begin{centering}
\includegraphics[width=0.5\columnwidth]{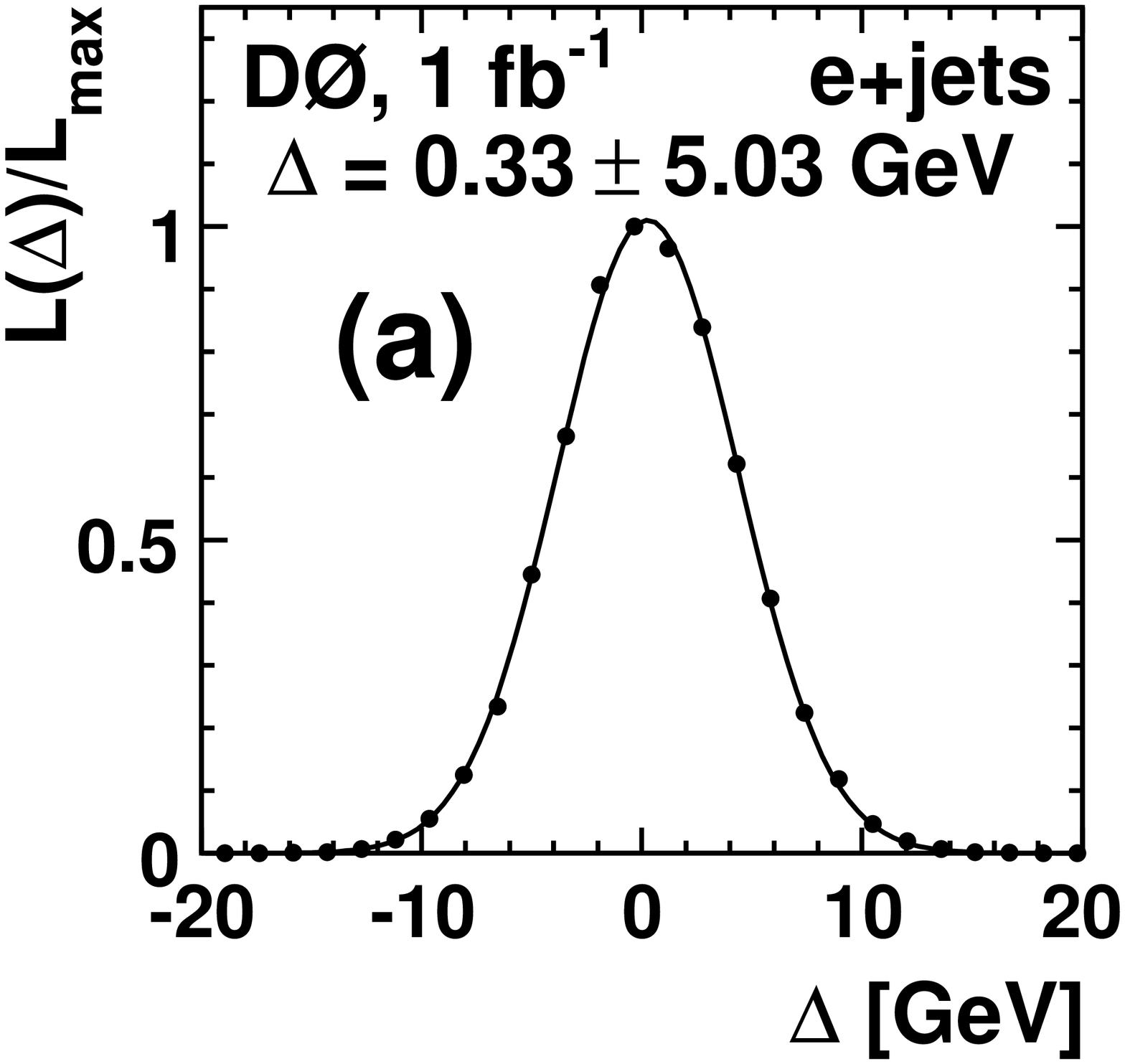}\includegraphics[width=0.5\columnwidth]{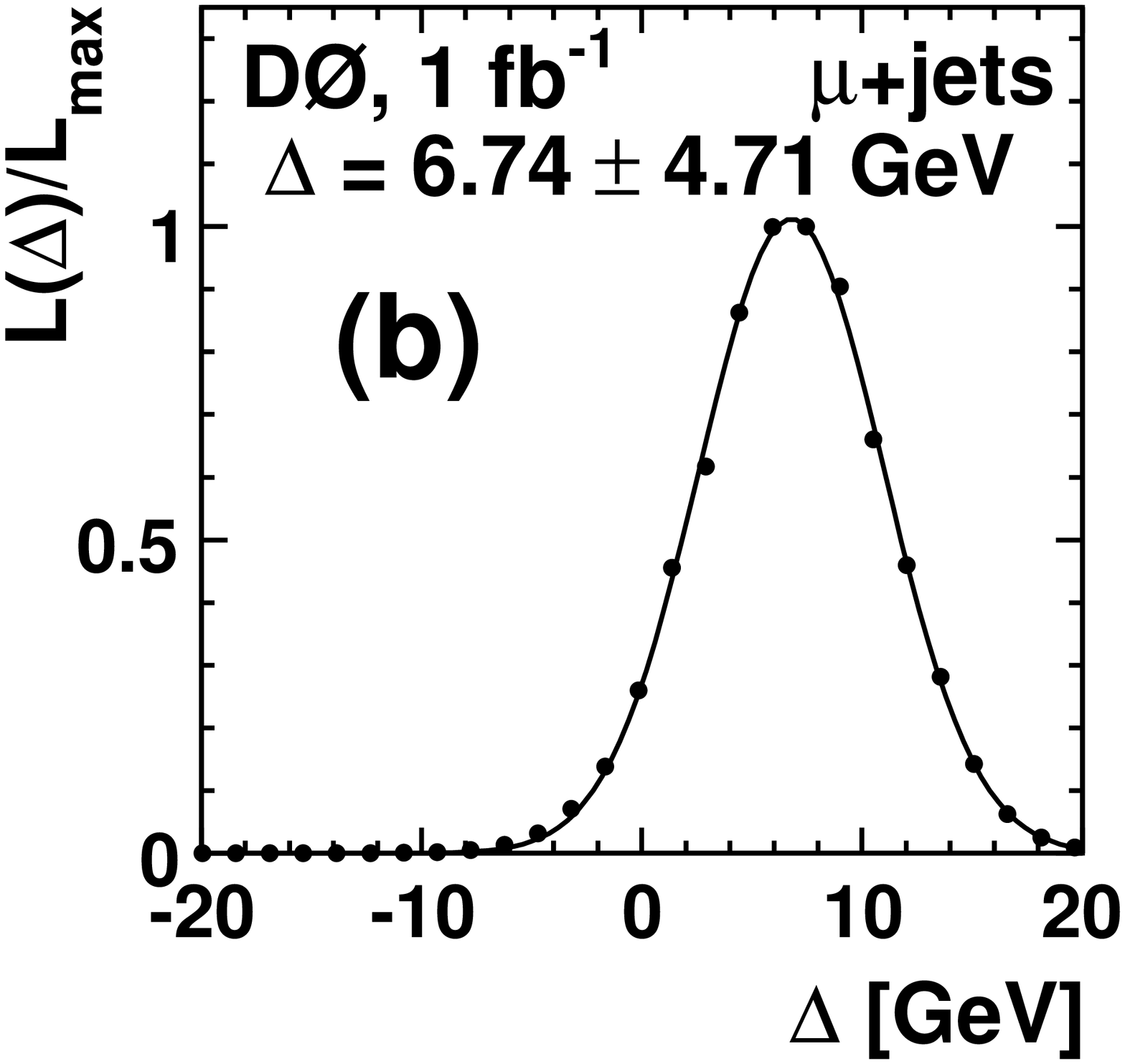} 
\par\end{centering}
\caption{\label{fig:mdelproj}Projections of the likelihoods onto the $\Delta$ axis
for (a) $e+$jets and (b) $\mu+$jets data.}
\end{figure}

Fitted two-dimensional Gaussian contours of equal probability (in terms of the
standard deviation sd) for $L(x;\,m_t,m_{\overline{t}})$ are  shown for the
electron and muon data samples in Figs.~\ref{fig:2d}(a) and \ref{fig:2d}(b), 
respectively. The corresponding $L(x;\Delta)$ for both channels are given in
Figs.~\ref{fig:mdelproj}(a) and \ref{fig:mdelproj}(b). The two sets of data are
consistent within their respective uncertainties, and the small correlations
($\rho^{e+{\rm jets}}=-0.05$, $\rho^{\mu+{\rm jets}}=-0.01$) extracted from the
fits in Fig.~\ref{fig:2d} between $m_t$ and $m_{\overline{t}}$ are not
statistically significant, nor are the shifts in the projections shown in
Fig.~\ref{fig:mdelproj}.

Results from the two channels are combined through a weighted average of the
separate electron and muon values. This has the advantage of  using their
respective pulls to adjust the uncertainties of each measurement before
combining the two results. Using this averaging process, we quote the final
combined means and their statistical uncertainties as
$\Delta=3.8\pm3.4(\textrm{stat.})$ GeV and $m_{\rm
sum}=170.9\pm1.5(\textrm{stat.})$ GeV.  The latter is consistent with the
previous measurement of Ref.~\cite{P17PRL} (see also Ref. \cite{noprior}).

The systematic uncertainties are summarized in Table~\ref{tab:syst}.  The first
category, {\it Physics modeling}, comprises the uncertainties in MC modeling of
$t\overline{t}$ and \wjets events.  The second category, {\it Detector modeling},
addresses uncertainties in the calibration of jet energy and  simulation of
detector response.  The last category, {\it Method}, addresses  uncertainties in the
calibration and possible systematic effects due to assumptions made in the
analysis. Except for two, all systematic uncertainties are identical to those 
described previously~\cite{P17PRL}.  Many of these uncertainties (e.g.,
uncertainties in JES, PDF, jet resolution, multijet contamination)
are expected to partially cancel in the measurement  of the
mass difference, but are often dominated by the statistics of the samples used
to evaluate them.  The two new contributions address the possibilities of (i)
reconstructing leptons with the wrong charge, and (ii) uncertainties from
modeling differences in the response of the calorimeter to $b$ and
$\overline{b}$ jets~\cite{bbar}, which can affect the measurement of the mass
difference.  These were evaluated for (i) by estimating the effect of an
increase in charge misidentification in MC simulations that would match that found in data
($\sim$1\% for both $e$ and $\mu$). For (ii), studies were performed on MC
samples and on data seeking any difference in detector response to $b$ and
$\overline{b}$ quarks beyond expectations from interactions of their decay
products, which are accommodated in the MC simulations. The observed differences were
limited by the statistics of both samples. The total systematic uncertainty is
1.2 GeV. Combining the systematic and statistical uncertainties of the
measurement in quadrature yields $\Delta=3.8\pm3.7$ GeV, a value consistent with
$CPT$ invariance.
\begin{table}
\caption{\label{tab:syst}Summary of systematic uncertainties on $\Delta$.}
\vspace{.25cm}
\centering
\begin{tabular}{lc}
\hline
\hline 
Source  & Uncertainty (GeV)\\
\hline
\multicolumn{2}{l}{\textit{Physics modeling}}\\
\hspace{12pt}Signal & $\pm0.85$\\
\hspace{12pt}PDF uncertainty      & $\pm0.26$\\
\hspace{12pt}Background modeling  & $\pm0.03$\\
\hspace{12pt}Heavy flavor scale factor   & $\pm0.07$\\
\hspace{12pt}$b$ fragmentation    & $\pm0.12$\\
\multicolumn{2}{l}{\textit{Detector modeling:}}\\
\hspace{12pt}$b$/light response ratio & $\pm0.04$\\
\hspace{12pt}Jet identification  & $\pm0.16$\\
\hspace{12pt}Jet resolution  & $\pm0.39$\\
\hspace{12pt}Trigger  & $\pm0.09$\\
\hspace{12pt}Overall jet energy scale & $\pm0.08$\\
\hspace{12pt}Residual jet energy scale & $\pm0.07$\\
\hspace{12pt}Muon resolution  & $\pm0.09$\\
\hspace{12pt}Wrong charge leptons & $\pm0.07$\\
\hspace{12pt}Asymmetry in $b\overline{b}$ response & $\pm0.42$\\
\multicolumn{2}{l}{\textit{Method:}}\\
\hspace{12pt}MC calibration & $\pm0.25$\\
\hspace{12pt}$b$-tagging efficiency  & $\pm0.25$\\
\hspace{12pt}Multijet contamination  & $\pm0.40$\\
\hspace{12pt}Signal fraction  & $\pm0.10$\\
Total (in quadrature)  & $\pm1.22$\\
\hline
\hline
\end{tabular}
\end{table}
 
In summary, we have measured the $t$ and $\overline{t}$ mass difference in 
$\sim$1 fb$^{-1}$ of data in $\ell+$jets $t\overline{t}$ events and find the
mass difference to be $m_t - m_{\overline{t}}=3.8\pm3.7$ GeV, corresponding to a
relative mass difference of $\Delta/m_{\rm sum}=(2.2\pm2.2)$\%.  This is the
first direct measurement  of a mass difference between a quark and its antiquark
partner.
 
%
We thank the staffs at Fermilab and collaborating institutions, 
and acknowledge support from the 
DOE and NSF (USA);
CEA and CNRS/IN2P3 (France);
FASI, Rosatom and RFBR (Russia);
CNPq, FAPERJ, FAPESP and FUNDUNESP (Brazil);
DAE and DST (India);
Colciencias (Colombia);
CONACyT (Mexico);
KRF and KOSEF (Korea);
CONICET and UBACyT (Argentina);
FOM (The Netherlands);
STFC and the Royal Society (United Kingdom);
MSMT and GACR (Czech Republic);
CRC Program, CFI, NSERC and WestGrid Project (Canada);
BMBF and DFG (Germany);
SFI (Ireland);
The Swedish Research Council (Sweden);
CAS and CNSF (China);
and the
Alexander von Humboldt Foundation (Germany).
%


\begin{thebibliography}{99}
\bibitem{pbkgqcd}
When this is satisfied, the relative contribution of $P_{\rm sig}$ to $P_{\rm
evt}$ will be negligible for such events, minimizing their influence in the
determination of  $m_{t}$ and $m_{\overline{t}}$.
\bibitem{nusol}
The transverse components of the unmeasured $\nu$ momentum $p_\nu$ are determined
from the $p_T$ balance of the \ttbar event, but the remaining ambiguity in the
longitudinal component of $p_\nu$ leaves more than one
possibility or ``solution'' for $p_\nu$.
\bibitem{sigtot}
$\sigma^{t\overline{t}}_{q\overline{q}}=\frac{8\pi\alpha_s^2}{27s^{5/2}}p(3E_t^2+3E_{\overline{t}}^2+2p^2+6m_tm_{\overline{t}})$
where $\alpha_s$ is the strong coupling constant, $s$ is the square of the
center-of-mass energy in the incoming $q\overline{q}$ rest frame, and $E_i$ and
$p$ are the energies of the $t$ or $\overline{t}$ and their common momentum,
respectively, in the $q\overline{q}$ frame.
\bibitem{noprior}
The result in Ref.~\cite{P17PRL} was obtained by further constraining the jet
energy scale to the one derived from photon+jets and dijet samples.  This
constraint is not used in the present analysis since it shifts $m_t$ and
$m_{\overline{t}}$ in the same way and has no effect on $\Delta$.
\bibitem{bbar}
Differences in the composition of $b$ and $\overline{b}$ jets, such as
different $K^+/K^-$ fractions, can cause differences in calorimeter response because
of differences in $K^+/K^-$ interaction cross sections.
\end{thebibliography}
\end{document}